# NEW EMBEDDING OF SCHWARZSCHILD GEOMETRY

# I. EXTERIOR SOLUTION

**Rainer Burghardt**[i]

We embed Schwarzschild geometry in a five-dimensional flat space by using two surfaces. Covariant field equations are deduced for the gravitational forces.

# 1. INTRODUCTION

It was pointed out by Eisenhart [1] that if any n-dimensional Ricci-flat metric can be embedded in an (n+1)-dimensional space, the metric is a flat metric. Thus, if the Schwarzschild metric is interpreted as the metric of a four-dimensional Riemannian space, one needs six dimensions at least for embedding. Kasner [2] proved this in particular for Einstein spaces and proposed for the Schwarzschild metric an immersion in six dimensions [3]. Since one of his co-ordinates is a transcendental integral, one co-ordinate line is missing in his theory. His embedding is a local embedding but not a global one [4].

If we rotate Schwarzschild's parabola with respect to its symmetry axis r through the angle $\Phi$, we can add the term $R^2 d\Phi^2$ to Kasner's line element. Looking at the transformation to cartesian co-ordinates, we find solutions to all integrals, but the extra components of the field equations one could derive from Kasner's theory can hardly be explained.

In order to avoid difficulties with the theorem of Eisenhart and Kasner, our new approach is to use *two* correlated four-dimensional surfaces embedded in a five-dimensional flat space. Then the Schwarzschild metric is the common metric and the Ricci tensor is the common Ricci tensor of both surfaces. This double surface may be interpreted as a single surface, the real world we are living in, with an additional function invoked by the second surface. It will be shown that the theory of a double surface is easily established. In chap. 2 we set up a general frame for spherically symmetric gravitation fields, in chap. 3 we specialize this frame to Schwarzschild geometry.

## 2. FIVE-DIMENSIONAL FORMULATION

Firstly, we develop a general five-dimensional formulation that could serve for four-dimensional Schwarzschild geometry and probably for other static spherically symmetric gravitation models. If the $X^{a'}$ are the components of a vector with value $X$ with respect to a cartesian co-ordinate system, a sphere of radius $X$ is parametrized in spherical polar co-ordinates by

$$\begin{aligned} X^{3'} &= X \sin\varepsilon \sin\vartheta \sin\varphi \\ X^{2'} &= X \sin\varepsilon \sin\vartheta \cos\varphi \\ X^{1'} &= X \sin\varepsilon \cos\vartheta \\ X^{0'} &= X \cos\varepsilon \sin i\psi \\ X^{4'} &= X \cos\varepsilon \cos i\psi \end{aligned} \qquad (2.1)$$

Differentiating (2.1), we find the tangent vectors to the spherical co-ordinate system

$$dX^a = \{dX^3, dX^2, dX^1, dX^0, dX^4\} = \{X\sin\varepsilon\sin\vartheta\, d\varphi, X\sin\varepsilon\, d\vartheta, Xd\varepsilon, dX, X\cos\varepsilon\, di\psi\} \qquad (2.2)$$

and the operators for partial differentiation $\hat{\partial}_a \Phi = \Phi_{,a}$, which are inverse to (2.2). In this co-ordinate system, $X^a$ has the only component $X^0 = X$. The line element in these co-ordinates is

$$dS^2 = X^2 \sin^2\varepsilon \sin^2\vartheta\, d\varphi^2 + X^2 \sin^2\varepsilon\, d\vartheta^2 + X^2 d\varepsilon^2 + dX^2 + X^2 \cos^2\varepsilon\, di\psi^2, \qquad (2.3)$$

which reduces to the line element on the sphere for $X = const$. We also could read off from (2.1) the transformation coefficients $D(i\psi, \varphi, \vartheta, \varepsilon)$ from cartesian to spherical co-ordinates, which are elements of the group SO(5). From the theory of coset spaces we get for our sphere $S^4$ = G/H=SO(5)/SO(4), where H is the group of transformations in the tangent space of $S^4$. The generator of the infinitesimal rotation in the [a,b]-plane is the operator $Q_{ab} = -Q_{ba}$. By differentiation of $D$ we find

$$\begin{aligned} D^{-1}dD &= e^{Q_{01}\varepsilon} e^{Q_{12}\vartheta} e^{Q_{23}\varphi} e^{Q_{04}i\psi} d e^{-Q_{04}i\psi} e^{-Q_{23}\varphi} e^{-Q_{12}\vartheta} e^{-Q_{01}\varepsilon} \\ &= Q_{10}\frac{1}{X}dX^1 + \left[Q_{20}\frac{1}{X} + Q_{21}\frac{1}{X}\cot\varepsilon\right]dX^2 \\ &\quad + \left[Q_{30}\frac{1}{X} + Q_{31}\frac{1}{X}\cot\varepsilon + Q_{32}\frac{1}{X\sin\varepsilon}\cot\vartheta\right]dX^3 + \left[Q_{40}\frac{1}{X} - Q_{41}\frac{1}{X}\tan\varepsilon\right]dX^4. \end{aligned} \qquad (2.4)$$

Defining the connection coefficients and covariant derivative in the spherical polar system by

$$X_{ab}{}^c = D_{c'}^c \hat{\partial}_a D_b^{c'}, \; X_a{}^{(bc)} = 0, \; \Phi_{a;b} = \Phi_{a,b} - X_{ba}{}^c \Phi_c, \qquad (2.5)$$

we can read off from (2.4) their components

$$X_{10}{}^1 = \frac{1}{X},\ X_{20}{}^2 = \frac{1}{X},\ X_{21}{}^2 = \frac{1}{X}\cot\varepsilon,$$
$$X_{30}{}^3 = \frac{1}{X},\ X_{31}{}^3 = \frac{1}{X}\cot\varepsilon,\ X_{32}{}^3 = \frac{1}{X\sin\varepsilon}\cot\vartheta, \tag{2.6}$$
$$X_{40}{}^4 = \frac{1}{X},\ X_{41}{}^4 = -\frac{1}{X}\tan\varepsilon.$$

The 4-bein $m_a = e_a^1, b_a = e_a^2, c_a = e_a^3, u_a = e_a^4$ defined on the sphere makes possible an alternative notation for the connection coefficients:

$$X_{ab}{}^c = \hat{M}_{ab}{}^c + \hat{B}_{ab}{}^c + \hat{C}_{ab}{}^c + \hat{U}_{ab}{}^c$$
$$\hat{M}_{ab}{}^c = m_a[\hat{M}_b m^c - m_b \hat{M}^c],\ \hat{B}_{ab}{}^c = b_a[\hat{B}_b b^c - b_b \hat{B}^c],\ \hat{C}_{ab}{}^c = c_a[\hat{C}_b c^c - c_b \hat{C}^c],\ \hat{U}_{ab}{}^c = u_a[\hat{U}_b u^c - u_b \hat{U}^c]. \tag{2.7}$$

The set of covariant derivatives

$$\hat{M}_{a;b} = \hat{M}_{a,b},\ \hat{B}_{a;b} = \hat{B}_{a,b} - \hat{M}_{ba}{}^c \hat{B}_c,\ \hat{C}_{a;b} = \hat{C}_{a,b} - \hat{M}_{ba}{}^c \hat{C}_c - \hat{B}_{ba}{}^c \hat{C}_c,\ \hat{U}_{a;b} = \hat{U}_{a,b} - \hat{M}_{ba}{}^c \hat{U}_c - \hat{B}_{ba}{}^c \hat{U}_c - \hat{C}_{ba}{}^c \hat{U}_c$$
$$\underset{1}{} \qquad \underset{2}{} \qquad \underset{3}{} \qquad \underset{4}{} \tag{2.8}$$

has the advantage that a tensor of an (n-m)-dimensional subspace covariantly differentiated by the (n-m)-method of (2.8) is also a tensor of the same (n-m)-dimensional subspace. The derivatives have the nice property

$$m_{a;b} = 0,\ b_{a;b} = 0,\ c_{a;b} = 0,\ u_{a;b} = 0\ . \tag{2.9}$$
$$\underset{1}{} \qquad \underset{2}{} \qquad \underset{3}{} \qquad \underset{4}{}$$

This method is a powerful tool for formulating covariantly field equations and (n-m)-dimensional Gaussian theorems. Inserting the relations above in the Ricci tensor

$$R_{ab} = X_{ab}{}^c{}_{,c} - X_{cb}{}^c{}_{,a} - X_{da}{}^c X_{cb}{}^d + X_{ab}{}^c X_{dc}{}^d \equiv 0, \tag{2.10}$$

the equations (2.7) decouple:

$$\hat{M}_{b;a} + \hat{M}_b \hat{M}_a = 0,\ \hat{B}_{b;a} + \hat{B}_b \hat{B}_a = 0,\ \hat{C}_{b;a} + \hat{C}_b \hat{C}_a = 0,\ \hat{U}_{b;a} + \hat{U}_b \hat{U}_a = 0,$$
$$\underset{1}{} \qquad \underset{2}{} \qquad \underset{3}{} \qquad \underset{4}{} \tag{2.11}$$
$$\hat{M}^c{}_{;c} + \hat{M}^c \hat{M}_c = 0,\ \hat{B}^c{}_{;c} + \hat{B}^c \hat{B}_c = 0,\ \hat{C}^c{}_{;c} + \hat{C}^c \hat{C}_c = 0,\ \hat{U}^c{}_{;c} + \hat{U}^c \hat{U}_c = 0,$$
$$\underset{1}{} \qquad \underset{2}{} \qquad \underset{3}{} \qquad \underset{4}{}$$

where

$$|\hat{M}_a| = \frac{1}{X},\ |\hat{B}_a| = \frac{1}{X\sin\varepsilon},\ |\hat{C}_a| = \frac{1}{X\sin\varepsilon\sin\vartheta},\ |\hat{U}_a| = \frac{1}{X\cos\varepsilon} \tag{2.12}$$

are the curvatures of several circular intersections of the sphere (the inverse of the values of the curvature vectors) and the equations (2.11) are the field equations for the curvatures of these intersections. They will constitute a general frame for Schwarzschild field equations and possibly for other static gravitational theories. The Einstein equations

$$G_{ab} = -\left[\hat{M}_{b;a} + (m_a m_b - g_{ab})\hat{M}^c{}_{;c} + {}^1t_{ab}\right]_1 - \left[\hat{B}_{b;a} + (b_a b_b - g_{ab})\hat{B}^c{}_{;c} + {}^2t_{ab}\right]_2$$
$$-\left[\hat{C}_{b;a} + (c_a c_b - g_{ab})\hat{C}^c{}_{;c} + {}^3t_{ab}\right]_3 - \left[\hat{U}_{b;a} + (u_a u_b - g_{ab})\hat{U}^c{}_{;c} + {}^4t_{ab}\right]_4 \equiv 0 \tag{2.13}$$

exhibit the conserved[1] quantities

$$\begin{aligned}
{}^1t_{ab} &= \hat{M}_a \hat{M}_b + (m_a m_b - g_{ab})\hat{M}^c \hat{M}_c, & {}^1t_a{}^b{}_{;b} &= 0, \\
{}^2t_{ab} &= \hat{B}_a \hat{B}_b + (b_a b_b - g_{ab})\hat{B}^c \hat{B}_c, & {}^2t_a{}^b{}_{;b} &= 0, \\
{}^3t_{ab} &= \hat{C}_a \hat{C}_b + (c_a c_b - g_{ab})\hat{C}^c \hat{C}_c, & {}^3t_a{}^b{}_{;b} &= 0, \\
{}^4t_{ab} &= \hat{U}_a \hat{U}_b + (u_a u_b - g_{ab})\hat{U}^c \hat{U}_c, & {}^4t_a{}^b{}_{;b} &= 0.
\end{aligned} \tag{2.14}$$

Cutting off the extra dimension (a, b=0), the Ricci tensor on the sphere $X = const.$ has the well known form

$$P_{mn} = \frac{3}{X^2} g_{mn}, \quad m,n = 1,...,4. \tag{2.15}$$

For later use we note

$$X_{,0} = 1, \ X_{,1} = 0, \ a = \cos\varepsilon, \ v = \sin\varepsilon, \ a_{,0} = 0, \ a_{,1} = -\frac{v}{X}, \ v_{,0} = 0, \ v_{,1} = \frac{a}{X}. \tag{2.16}$$

## 3. THE FOUR-DIMENSIONAL THEORY

We expand the relation $X_{a'} X^{a'} = X^2$ to

$$(x_{a'} - \bar{x}_{a'})(x^{a'} - \bar{x}^{a'}) = X^2, \tag{3.1}$$

where the $x^{a'}$ are the co-ordinates of the tip and the $\bar{x}^{a'}$ those of the origin of the vector $X^{a'}$. To begin with, we reduce the dimensions of our problem to $\{x^{0'}, x^{1'}\} = \{R, r\}, \{\bar{x}^{0'}, \bar{x}^{1'}\} = \{\bar{R}, \bar{r}\}$. Then it is easy to show that Schwarzschild's parabola and Neil's parabola[2] (evolute of Schwarzschild's parabola)

---

[1] There does not seem to be any relation to physical conservation laws.
[2] We only use the positive and negative branch respectively.

$$R^2 = 8M(r-2M), \quad \overline{R}^2 = \frac{2}{M}\left(\frac{\overline{r}}{3} - 2M\right)^3 \qquad (3.2a,b)$$

satisfy (3.1), if $X = \sqrt{2r^3/M}$ is the value of the vector of curvature $X^{a'}$ of the Schwarzschild parabola and the relation $\overline{r} = 3r$ is used. Since all curves parallel to Schwarzschild's parabola with respect to their normals have the same evolute (3.2b), equ. (3.1) is satisfied by this family of curves and Neil's parabola. These curves and their orthogonal trajectories, which are straight lines and tangents to Neil's parabola set up a co-ordinate system in the [0',1']-plane. Rotating these curves through the angles $\vartheta, \varphi, i\psi$, an all-over-space curvilinear co-ordinate system suitable for describing Schwarzschild theory is fixed. The Schwarzschild parabola by rotation through $\vartheta$ and $\varphi$ creates Flamm's hyper-paraboloid [5]. The real image of the imaginary rotation through $i\psi$ of (3.2a) is a hyperbolic-parabolic surface. Neil's parabola is rotated in the same way and generates the second surface. For the description of the double-surface theory we use the variables $\{r, R, \overline{r}, \overline{R}, \vartheta, \varphi, i\psi\}$. Since the evolute is the same for the Schwarzschild parabola and all curves parallel to it, one variable is redundant and can be eliminated by (3.2.b). Therefore, we are left with *six* variables $\{r, R, \overline{r}, \vartheta, \varphi, i\psi\}$ in agreement with Kasner's proof [2]. The difference to Kasner is that three of them are necessary to describe the geometry in the [0',1']-plane. We will see that the spherical structure of group space discussed in chap. 1 and its map to a parabolic double surface provide the possibility for embedding Schwarzschild geometry in a five-dimensional flat space.

Restricting ourselves to two dimensions again and rewriting (3.1) and (2.1) as

$$(R-\overline{R})^2 + (r-\overline{r})^2 = X^2, \; X = X(r,\overline{r},R,\overline{R}), \; X^{0'} = X\cos\varepsilon = Xa, \; X^{1'} = X\sin\varepsilon = Xv, \qquad (3.3)$$

we extract the all-over-space functions

$$a(r,\overline{r},R,\overline{R}) = \frac{R-\overline{R}}{X}, \quad v(r,\overline{r},R,\overline{R}) = \frac{r-\overline{r}}{X}, \quad a^2 + v^2 = 1. \qquad (3.4)$$

Evidently, these quantities are constant along the straight lines of the co-ordinate net but vary along the parallel curves. Defining the derivatives with respect to local reference systems spanned by the co-ordinates defined above

$$\partial_0 = a\frac{\partial}{\partial R} + v\frac{\partial}{\partial r}, \; \partial_1 = -v\frac{\partial}{\partial R} + a\frac{\partial}{\partial r}, \; \Phi_{|a} = \partial_a \Phi,$$
$$\partial_{\overline{0}} = a\frac{\partial}{\partial \overline{R}} + v\frac{\partial}{\partial \overline{r}}, \; \partial_{\overline{1}} = -v\frac{\partial}{\partial \overline{R}} + a\frac{\partial}{\partial \overline{r}}, \; \Phi_{|\overline{a}} = \partial_{\overline{a}} \Phi, \qquad (3.5)$$

and applying this on (3.4), we get

$$a_{|0} = 0, \; a_{|1} = -\frac{v}{X}, \; v_{|0} = 0, \; v_{|1} = \frac{a}{X}, \; a_{|\overline{0}} = 0, \; a_{|\overline{1}} = \frac{v}{X}, \; v_{|\overline{0}} = 0, \; v_{|\overline{1}} = -\frac{a}{X}. \qquad (3.6)$$

We note, that the slope of the parabola and its parallel curves is $-\tan\varepsilon = -v/a$, where $\varepsilon$ is taken to be CW for later convenience and has the range [π/2, 0]. $v$ is negative for the positive branches of these curves. For the straight lines with ascent $dR/dr = a/v$ and for the curves with ascent $dR/dr = -v/a$ the tangent vectors are

$$dx^0 = adR + vdr = \frac{1}{v}dr, \quad dx^1 = -vdR + adr = \frac{1}{a}dr, \qquad (3.7a)$$

and for Neil's parabola with ascent $d\overline{R}/d\overline{r} = a/v$

$$d\overline{x}^0 = ad\overline{R} + vd\overline{R} = \frac{1}{v}d\overline{r}, \quad d\overline{x}^1 = -vd\overline{R} + ad\overline{r} = 0. \qquad (3.7b)$$

The change of the vector of curvature

$$dX^a = \{dx^0 - d\overline{x}^0, dx^1\} \qquad (3.8)$$

has three contributions: $dx^0$ by proceeding from a curve to a neighboring curve, $d\overline{x}^0$ on the evolute and $dx^1$ on the parallel curves of the net by constraining the motion of $X^a$ to one of these curves.

Considering the Schwarzschild parabola (3.2a) and applying the relations below (3.2), we get, from (3.4),

$$a = \sqrt{1 - 2M/r}, \quad v = -\sqrt{2M/r}, \qquad (3.9)$$

where $v$ is the velocity of a freely falling observer and $a^{-1}$ the Lorentz factor of this motion. The vector of curvature constrained to Schwarzschild parabola is

$$\rho(r) = X\big|_{parabola} = \sqrt{2r^3/M}, \quad \partial_1\rho(r) = -3\frac{a}{v}, \qquad (3.10a,b)$$

where (3.10b) deserves some attention. In contrast to the all-over-space function $X(r,R,\overline{r},\overline{R})$, $\partial_1 X = 0$, the changes (3.10b) of $\rho(r)$ can be measured on the Schwarzschild parabola. For better understanding we interpret $\rho(r)$ as the parameter of the family of orthogonal trajectories of the parabola, where $\rho = const.$ denotes a specific line of this family. Therefore, we can take $\rho(r)$ as a potential and $\partial_1\rho(r)$ as the gradient of this potential. Thus $\rho(r)$ is a well-defined function on the Schwarzschild parabola. With the aid of (3.3), (3.4) we are able to split the space-like tangent vectors of the group space into the tangent vectors of the two surfaces of rotation

$$dX^2 = Xvd\vartheta = rd\vartheta - \overline{r}d\vartheta, \quad dX^3 = Xv\sin\vartheta d\varphi = r\sin\vartheta d\varphi - \overline{r}\sin\vartheta d\varphi. \qquad (3.11)$$

The time-like vector results from two translation surfaces with a topology different from those of the space-like surfaces of rotation. The Schwarzschild co-ordinate time element is the infinitely small area

$$dt = \rho\, d\psi \qquad (3.12a)$$

arising from the transport of $\rho$ along the hyperbolic intersections of the translation surfaces by $d\psi$. The physical time element

$$dX^4 = i\rho \, ad\psi = iadt \tag{3.12b}$$

is the projection of this area onto the plane normal to the symmetry axis of the parabola.

With the definition of the projector (the upper index refers to the group space and the lower index to the double surface)

$$\mathrm{p}_3^3 = \mathrm{p}_2^2 = \frac{Xv}{r}, \quad \mathrm{p} = \mathrm{p}_0^0 = \mathrm{p}_1^1 = \mathrm{p}_4^4 = \frac{X}{\rho}, \quad \mathrm{p}_2^2\Big|_{parabola} = -2, \quad \mathrm{p}\big|_{parabola} = 1, \tag{3.12}$$

we are able to calculate the induced metric (m, n =1,2,...,4)

$$dx^c = (\mathrm{p}^{-1})_a^c dX^a, \quad ds^2 = g_{mn}(\mathrm{p}^{-1})_a^m (\mathrm{p}^{-1})_b^n dX^a dX^b = \rho^2 d\varepsilon^2 + r^2 d\vartheta^2 + r^2 \sin^2\vartheta \, d\varphi^2 - \rho^2 a^2 d\psi^2. \tag{3.13}$$

As we have to take for the parabola the constrained functions (3.9,3.10a), we get with $dv = ad\varepsilon$, $dv(r) = (1/\rho(r))\, dr$ and $\rho(r)d\varepsilon = (1/a(r))\, dr$ the Schwarzschild metric

$$ds^2 = a^{-2}dr^2 + r^2 d\vartheta^2 + r^2 \sin^2 d\varphi^2 - a^2 dt^2, \; a = \sqrt{1 - 2M/r}. \tag{3.14}$$

We note that (3.14) is *not* the metric on the surface created by rotation of the Schwarzschild parabola but the metric of this surface with an additional function $\rho(r)$ invoked by the second surface. This we call the physical surface.

## 4. THE FIELD EQUATIONS

By use of the projectors (3.13), we derive the connection coefficients and the covariant derivative due to the displacement $dx^a$

$$\partial_a = \mathrm{p}_a^b \hat{\partial}_b, \quad A_{ab}{}^c = \mathrm{p}_a^d X_{db}{}^c, \quad \Phi_{a\|b} = \Phi_{a|b} - A_{ba}{}^c \Phi_c \tag{4.1}$$

and we recover the projectors (3.12) by $X^a{}_{\|b} = \mathrm{p}_b^a$ in contrast to $X^a{}_{;b} = \delta_b^a$, $X^{a'}{}_{;b} = D_b^{a'}$. From (2.6), we get new quantities

$$M_0 = \frac{1}{\rho}, B_0 = \frac{v}{r}, B_1 = \frac{a}{r}, C_0 = \frac{v}{r}, C_1 = \frac{a}{r}, C_2 = \frac{1}{r}\cot\vartheta, U_0 = \frac{1}{\rho}, U_1 = -\frac{1}{\rho}\frac{v}{a} \tag{4.2}$$

and a set equivalent to (2.7). Inserting (3.10a) into $U_1$, we get for $-U_1$ the Schwarzschild gravitational field strength $-\frac{1}{a}\frac{M}{r^2}$. We obtain the projection of the Riemann tensor by

$$R_{abc}{}^d(A) = 2\left[ A_{[b \cdot c \cdot |a]}{}^d + A_{[b \cdot c \cdot}{}^f A_{a]f}{}^d + A_{[ba]}{}^f A_{fc}{}^d \right],$$
$$\mathrm{p}_a^g \mathrm{p}_b^h R_{ghc}{}^d(X) = R_{abc}{}^d(A) + 2X_{fc}{}^d \mathrm{p}_{[a\|b]}^f = 0. \tag{4.3}$$

We note that $R_{abc}{}^d(A)$ is the curvature tensor of the physical surface defined above, but *not* the Riemann tensor of one single member of the double surface discussed in chap. 3. But we also note that the description of a double surface is fairly equivalent to the description of a single surface. Thus it is not surprising that gravitational field equations may be interpreted in the framework of a double surface theory. Evaluating the last term of (4.3), we have to pay attention to the case where the motion of $\rho$ is constrained to the parabola. The action of

$$\mathrm{p} = \frac{X(r,R,\bar{r},\bar{R})}{\rho(r)}$$

on a quantity is to interchange the functional dependence. Differentiating $\mathrm{p}$ we then have to use (3.10b). Instead of (2.11), we get, from the contraction of (4.3), the field equations

$$M_{a\|b} + M_a M_b = M_a \mathrm{p}_{,b}, \quad B_{a\|b} + B_a B_b = 0, \quad C_{a\|b} + C_a C_b = 0, \quad U_{a\|b} + U_a U_b = U_a \mathrm{p}_{,b},$$
$$\underset{1}{\phantom{M}} \qquad \underset{2}{\phantom{B}} \qquad \underset{3}{\phantom{C}} \qquad \underset{4}{\phantom{U}}$$
(4.4)
$$\mathrm{p}_{,0} = 0, \quad \mathrm{p}_{,1} = 3\rho \frac{a}{v} \ .$$

If all terms with summation over the extra dimension are isolated, the four-dimensional reduction of (4.3) can be written as

$$R_{mnr}{}^s(A) + Z_{mnr}{}^s = 0, \quad Z_{mnr}{}^s = A_{nr}{}^0 A_{m0}{}^s - A_{mr}{}^0 A_{n0}{}^s + 2 X_{tr}{}^s \mathrm{p}^t_{[m\|n]},$$
(4.5)

where the extra term Z supplies the components

$$Z_{121}{}^2 = Z_{131}{}^3 = Z_{242}{}^4 = Z_{343}{}^4 = -\frac{2}{\rho^2}, \quad Z_{141}{}^4 = Z_{232}{}^3 = \frac{4}{\rho^2}$$

expected for the Schwarzschild theory, and $R_{mns}{}^r(A)$ has Riemannian properties. As $Z_{rmn}{}^r = 0$ the embedded physical surface is Ricci-flat. In four dimensions (4.4) is less amazing, though there are more expressions on the right side of these equations, having their origin from the extra dimensions. Therefore we restrict ourselves to noting the Schwarzschild Ricci tensor in a covariant form[3]

$$R_{mn}(A) = -\left[ B_{m\|n} + B_m B_n \right] - b_m b_n \left[ B^r{}_{\|r} + B^r B_r \right] - \left[ C_{m\|n} + C_m C_n \right] - c_m c_n \left[ C^r{}_{\|r} + C^r C_r \right]$$
$$\underset{2}{\phantom{B}} \qquad \underset{2}{\phantom{B}} \qquad \underset{3}{\phantom{C}} \qquad \underset{3}{\phantom{C}}$$
$$-\left[ U_{m\|n} + U_m U_n \right] - u_m u_n \left[ U^r{}_{\|r} + U^r U_r \right] = 0, \quad B_{[m\|n]} = 0, C_{[m\|n]} = 0, U_{[m\|n]} = 0.$$
$$\underset{4}{\phantom{U}} \qquad \underset{4}{\phantom{U}} \qquad \underset{2}{\phantom{B}} \qquad \underset{3}{\phantom{C}} \qquad \underset{4}{\phantom{U}}$$
(4.6)

---

[3] If four-dimensional labels m,n,r are combined with the operator for covariant differentiation than only the four-dimensional components of the connection coefficients are used.

The Schwarzschild gravitational field strength $g = -U_1$ differs from Newton's analogue $g_N = -km/r^2$ by the factor $\alpha = a^{-1} = 1/\sqrt{1-2M/r}$ and is increasing towards the center of gravitation stronger than $g_N$ and tends towards an infinite value at $r = 2M$. As the last bracket of the field equation vanishes, we have

$$div g = g^2, \qquad (4.7)$$

where $g^2$ is the energy density due to the self-interaction [7,8] of the gravitational field. By integration over the volume $\int_V dV = \int_{r_s}^{\infty}\int_0^{\pi}\int_0^{2\pi} \alpha\, r^2 \sin\vartheta\, dr\, d\vartheta\, d\varphi$, $r_s$ being the radial Schwarzschild co-ordinate for the surface of a stellar object, we get

$$\int_V div g\, dV = 4\pi km[\alpha(r_s) - 1]. \qquad (4.8)$$

If we compare (4.5) with the corresponding Newtonian expression $\int div g_N dV_N = -4\pi km$ we find the field energy to be repulsive, as a negative mass would be. The co-ordinate r of the flat embedding space is not suitable for describing the radial line element $dx^1 = (1/a)dr$ at the vertex $r = 2M$ of the Schwarzschild parabola. With the slope $dR/dr = -v/a$ we also get $dx^1 = -(1/v)dR$. At the vertex the tangent vector $dx^1 = dR$ is normal to the symmetry axis of the parabola. The physical surface is regular everywhere and there is not much room for time journeys and all that in our purist geometrical picture. As M is the parameter for the mass of a stellar object covering a region from $r = 0$ to $r_s$, the exterior Schwarzschild solution has to be extended beyond $r_s$ by the interior solution. The geometrical properties of the space-like part of the interior solution has been elucidated by Flamm, 1916 [5]. The space-like interior metric is the metric of a sphere with radius $\mathcal{R}$. One easily finds a relation of this radius to the Schwarzschild parabola: $\rho(r_s) = 2\mathcal{R}$. Since any realistic model of a stellar object has to be described by a complete solution, the Schwarzschild throat is covered by a trough, and problems with 'singularities' don't occur. Treder et al. [6] considered a general ansatz for a time-dependent Schwarzschild solution which reduces via Birkhoff-theorem to the complete static Schwarzschild solution.

## 5. THE FREELY FALLING SYSTEM

We envisage now an interesting example of the action of the group H, the group of transformations in the tangent space of the sphere $S^4$. The transition to a freely falling system (in this chapter denoted by primed indices) is performed by the transformation

$$H_1^{1'} = H_4^{4'} = \cos i\chi = \alpha,\ H_4^{1'} = -H_1^{4'} = \sin i\chi = i\alpha v,\ \alpha = \frac{1}{\sqrt{1-\frac{2M}{r}}},\ v = -\sqrt{\frac{2M}{r}}, \qquad (5.1)$$

which is a generalized Lorentz transformation if referred to the physical surface. The covariant differentiation in terms of the freely falling system includes the inhomogeneous term $\Lambda$:

$$\Phi_{a';b'} = H_{a'}^{a} H_{b'}^{b} \Phi_{a;b} = \Phi_{a',b'} - \Lambda_{b'a'}{}^{c'} \Phi_{c'} - X_{b'a'}{}^{c'} \Phi_{c'}, \quad \Lambda_{b'a'}{}^{c'} = H_{c}^{c'} H_{a',b'}^{c} = i\chi_{a',b'}^{c}, \quad \chi^{4'}{}_{1'} = \chi^{4}{}_{1} = \chi. \qquad (5.2)$$

but it would be rash to combine this inhomogeneous term with the connection coefficients, as is usually done in tensor analysis. One could show that with the transformed quantities

$$\hat{M}_{0'} = \hat{M}_0, \; \hat{B}_{0'} = \frac{1}{X}, \; \hat{B}_{1'} = \frac{1}{Xv}, \; \hat{B}_{4'} = -\frac{i}{X}, \; \hat{U}_{0'} = \hat{U}_0, \; \hat{U}_{1'} = \alpha \hat{U}_1, \; \hat{U}_{4'} = -i\alpha v \hat{U}_1 \qquad (5.3)$$

and with the definitions

$$\hat{M}_{\underset{1}{a';b'}} = \hat{M}_{a',b'} - \Lambda_{b'a'}{}^{c'} \hat{M}_{c'}, \hat{B}_{\underset{2}{a';b'}} = \hat{B}_{a',b'} - \Lambda_{b'a'}{}^{c'} \hat{B}_{c'} - \hat{M}_{b'a'}{}^{c'} \hat{B}_{c'},$$

$$\hat{C}_{\underset{3}{a';b'}} = \hat{C}_{a',b'} - \Lambda_{b'a'}{}^{c'} \hat{C}_{c'} - \hat{M}_{b'a'}{}^{c'} \hat{C}_{c'} - \hat{B}_{b'a'}{}^{c'} \hat{C}_{c'}, \qquad (5.4)$$

$$\hat{U}_{\underset{4}{a';b'}} = \hat{U}_{a',b'} - \Lambda_{b'a'}{}^{c'} \hat{U}_{c'} - \hat{M}_{b'a'}{}^{c'} \hat{U}_{c'} - \hat{B}_{b'a'}{}^{c'} \hat{U}_{c'} - \hat{C}_{b'a'}{}^{c'} \hat{U}_{c'}$$

the system of field equations (2.11-2.14) is invariant under the transformation of H. As the projectors $\mathfrak{p}_b^a$ are invariant too, the same holds for the system of equations (4.4) and (4.6). The primed field equations just stand for the predictions a freely falling observer would make for the physics of a static observer. The constituents of the field equations transform like tensors under H, but this is not surprising. The geometrical background of the quantities $\hat{M}, \hat{B}, \hat{C}, \hat{U}$ is the curvatures (2.12) of several intersections of the surface which are expected to be invariant.

Now we have to investigate how to formulate the physics of the freely falling observer. For this purpose, we rearrange some terms in the primed version of the Ricci tensor. We will show this only in short. As under H the bein vectors combine to

$$'m_{a'} = H_1^{1'} m_{a'} + H_4^{1'} u_{a'} = \delta_{a'}^{1'}, \; 'u_{a'} = H_1^{4'} m_{a'} + H_4^{4'} u_{a'} = \delta_{a'}^{4'},$$

the components of $\hat{M}$ and $\hat{U}$ get mixed under this transformation

$$'\hat{M}_{a'b'}{}^{c'} + '\hat{U}_{a'b'}{}^{c'} = H_{a'}^{a} H_{b'}^{b} H_{c}^{c'} [\hat{M}_{ab}{}^{c} + \hat{U}_{ab}{}^{c}] + \Lambda_{a'b'}{}^{c'}$$

$$'\hat{B}_{a'b'}{}^{c'} = H_{a'}^{a} H_{b'}^{b} H_{c}^{c'} \hat{B}_{ab}{}^{c}, \; '\hat{C}_{a'b'}{}^{c'} = H_{a'}^{a} H_{b'}^{b} H_{c}^{c'} \hat{C}_{ab}{}^{c} \qquad (5.5)$$

We now have made use of the inhomogeneous transformation in the conventional sense, but we still have to analyze this procedure in greater detail. Starting with (2.4), we reformulate the connection coefficients in such a manner that they are expressed by the covariant change of the angles

$$\varepsilon^{10} = \varepsilon, \vartheta^{20} = \vartheta \sin \varepsilon, \vartheta^{21} = \vartheta \cos \varepsilon,$$

$$\varphi^{30} = \varphi \sin \varepsilon \sin \vartheta, \varphi^{31} = \varphi \cos \varepsilon \sin \vartheta, \varphi^{32} = \varphi \cos \vartheta, \psi^{40} = \psi \cos \varepsilon, \psi^{41} = \psi \sin \varepsilon,$$

$$\hat{M}_{ab}{}^c = \varepsilon^c{}_{b;a} , \hat{B}_{ab}{}^c = \vartheta^c{}_{b;a} , \hat{C}_{ab}{}^c = \varphi^c{}_{b;a} , \hat{U}_{ab}{}^c = i\psi^c{}_{b;a} . \underset{1\phantom{xx}2\phantom{xx}3\phantom{xx}4}{} \tag{5.6}$$

Evidently, any combination of one of these quantities with $\Lambda_{a'b'}{}^{c'} = \chi^{c'}{}_{b',a'}$ indicates the change of two angles. Therefore, we conclude that the inhomogeneous transformation of a connection coefficient is a rule of how to generate from a geometrical object a *new* object with new geometrical properties. This idea is important for the physical interpretation of these objects as field strengths. If $U_{ab}{}^c$ is the field strength for a static observer then $'U_{a'b'}{}^{c'}$ is the field strength a relatively accelerated observer would experience in terms of his proper system. The transformation (5.5) results in

$$'\hat{M}_{0'} = \frac{1}{X}, \, '\hat{M}_{4'} = -\frac{i}{X}, \, '\hat{B}_{a'} = \hat{B}_{a'}, \, '\hat{C}_{a'} = \hat{C}_{a'}, \, '\hat{U}_{0'} = \frac{1}{X}. \tag{5.7}$$

We now drop all primes for the rest of the chapter. The new set of covariant derivatives is

$$\hat{M}_{a;b} = \hat{M}_{a,b} - \hat{U}_{ba}{}^c \hat{M}_c , \, \hat{B}_{a;b} = \hat{B}_{a,b} - \hat{M}_{ba}{}^c \hat{B}_c - \hat{U}_{ba}{}^c \hat{B}_c ,$$
$$\hat{C}_{a;b} = \hat{C}_{a,b} - \hat{M}_{ba}{}^c \hat{C}_c - \hat{B}_{ba}{}^c \hat{C}_c - \hat{U}_{ba}{}^c \hat{C}_c , \, \hat{U}_{a;b} = \hat{U}_{a,b} . \tag{5.8}$$

The evaluation of the Ricci tensor leads us to the same set of equations (2.11). The five-dimensional field equations are form invariant, if a transformation of the reference system in the tangent space and a correlated inhomogeneous transformation are simultaneously performed. The projected quantities

$$M_a = \left\{\frac{1}{\rho},0,0,0,-\frac{i}{\rho}\right\}, B_a = \left\{\frac{v}{r},\frac{1}{r},0,0,-\frac{iv}{r}\right\}, C_a = \left\{\frac{v}{r},\frac{1}{r},\frac{1}{r}\cot\vartheta,0,-\frac{iv}{r}\right\}, U_a = \left\{\frac{1}{\rho},0,0,0\right\} \tag{5.9}$$

with $a = 0,1,...,4$ satisfy (4.4), since $\partial_1 = \frac{\partial}{\partial r}$, $\partial_4 = -iv\frac{\partial}{\partial r}$, $p_0 = 0$, $p_1 = 3\frac{1}{\rho v}$, $p_4 = -3\frac{i}{\rho}$. No gravitational field strength occurs in (5.9) but there are the tidal forces

$$M_4 = \frac{iv}{2r}, B_4 = -\frac{iv}{r}, C_4 = -\frac{iv}{r}, \tag{5.10}$$

stretching and squeezing a falling object. From an earlier stage of development we get from (4.3) instead of (4.6) the field equations for the static system

$$R_{mn}(A) = -\left[M_{n\|m} - m_m M_{n\|r} m^r + M_m M_n\right] - m_m m_n \left[M^r{}_{\|r} + M^r M_r\right] - \left[B_{n\|m} + B_m B_n\right] - b_m b_n \left[B^r{}_{\|r} + B^r B_r\right]$$
$$\underset{1}{} \phantom{x} \underset{1}{} \phantom{xxxxxxxxx} \underset{2}{} \phantom{xxxxxxx} \underset{2}{}$$
$$-\left[C_{n\|m} + C_m C_n\right] - c_m c_n \left[C^r{}_{\|r} + C^r C_r\right] - \left[U_{n\|m} + U_m U_n\right] - u_m u_n \left[U^r{}_{\|r} + U^r U_r\right] = 0,$$
$$\underset{3}{} \phantom{xxxxxxx} \underset{3}{} \phantom{xxxxxxx} \underset{4}{} \phantom{xxxxxxx} \underset{4}{} \tag{5.11}$$

where the first two brackets are empty. For the freely falling system we get the same equations, but the last two brackets are empty. The four-dimensional field equations have the same invariance properties as the five-dimensional ones. We recover a theory for freely falling systems in Schwarzschild geometry, developed by us in an earlier paper [9].

# CONCLUSIONS

Although most people are of opinion that six dimensions are a presupposition for embedding Schwarzschild geometry, we have been able to show that five dimensions are sufficient for embedding. There are no contradictions to the proofs of Kasner and Eisenhart, if we make use of two correlated embedded surfaces, which demand the introduction of *six* variables. Amazingly, it is rather easy to deduce a geometrical theory for a double surface, because this double surface is a map of a single spherical surface of the group space. The Schwarzschild Ricci tensor refers to both surfaces of the double surface system and vanishes without demanding the Riemann tensor to be flat. Moreover, the theory provides a fully covariant representation of the Schwarzschild field equations. They are co-ordinate invariant, Lorentz invariant and form invariant.

# ACKNOWLEDGEMENTS

My thanks are due to Prof. Dr. H.-J. Treder for his kind support and for encouraging me, in numerous letters, to write this paper, D.I. Dr. P. Helnwein for fruitful discussions, and Mag. J. Johann for drawing my attention to the hyperbolic-parabolic surface.

# REFERENCES

1. L. P. Eisenhart, *Riemannian spaces*, Princeton 1925
2. E. Kasner, *Am. Journ. Math.* **43**, 126 (1921)
3. E. Kasner, *Am. Journ. Math.* **43**, 130 (1921)
4. H. -J. Treder, private communication.
5. L. Flamm, *Phys. Z.* **17**, 448 (1916)
6. Mercier, Treder, Yourgrau, *On general relativity and gravitation*, Berlin 1979
7. H.-J. Treder, *Ann. d. Phys.* **35**, 137 (1978)
8. H.-J. Treder and W. Yourgrau, *Phys. Lett.* **64A**, 25 (1977)
9. R. Burghardt, *Found. Phys. Lett.* **8**, 575 (1995)

---

[i] A-2943 Obritz 246, homepage: arg.at.tf, e-mail: arg@i-one.at